\documentclass[aps,prl,twocolumn,floatfix]{revtex4}
\usepackage{amsmath}
\usepackage{url}
\usepackage{subfigure}
\usepackage{graphicx,psfrag}
\usepackage{bm}
\usepackage{multirow}

\begin{document}

\title{Experimental Evidence for Mixed Reality States}
\author{Alfred H{\"u}bler}
\email{hubler.alfred@gmail.com}
\author{Vadas Gintautas}
\email{vadasg@gmail.com}
\altaffiliation[Also at the ]{Center for Nonlinear Studies, Theoretical Division, MS B284, Los Alamos National Laboratory, Los Alamos NM 87545, USA}
\affiliation{Center for Complex Systems Research, Dept. of Physics, University of Illinois at Urbana-Champaign, Urbana IL 61801, USA }
 
 \date{\today}

\maketitle

Recently researchers at the University of Illinois coupled a real pendulum to its virtual counterpart. They observed that the two pendulums suddenly start to move in synchrony if their lengths are sufficiently close~\cite{Gintautas07PRE}. In this synchronized state, the boundary between the real system and the virtual system is blurred, that is, the pendulums are in a mixed reality state. An instantaneous, bidirectional coupling is a prerequisite for mixed reality states. In the experimental setup, a motor moves the pivot point of the real pendulum slightly whenever the virtual pendulum moves, and vice versa. This requires a fast probe, a fast motor controller, and a fast computer that can run the virtual pendulum in real time and accomplish the communication between the probe and the motor controller.

When a real system is coupled to its virtual counterpart with a small, bidirectional, instantaneous coupling, the coupled system is an interreality system. A phase transition separates the two states of an interreality system: a dual reality state and a mixed reality state.

In a mixed reality state, (i) the motion of the real system and the motion of the virtual system are in step and synchronized; (ii) the energy of the entire system is conserved; (iii) the amplitude of the motion increases to a large limiting amplitude; (iv) the real system strengthens the motion of the virtual system and vice versa; and thus (v) the interreality system can generate a large output signal.

\begin{figure}[htb]
    \begin{center}
      \includegraphics[width=0.98\columnwidth]{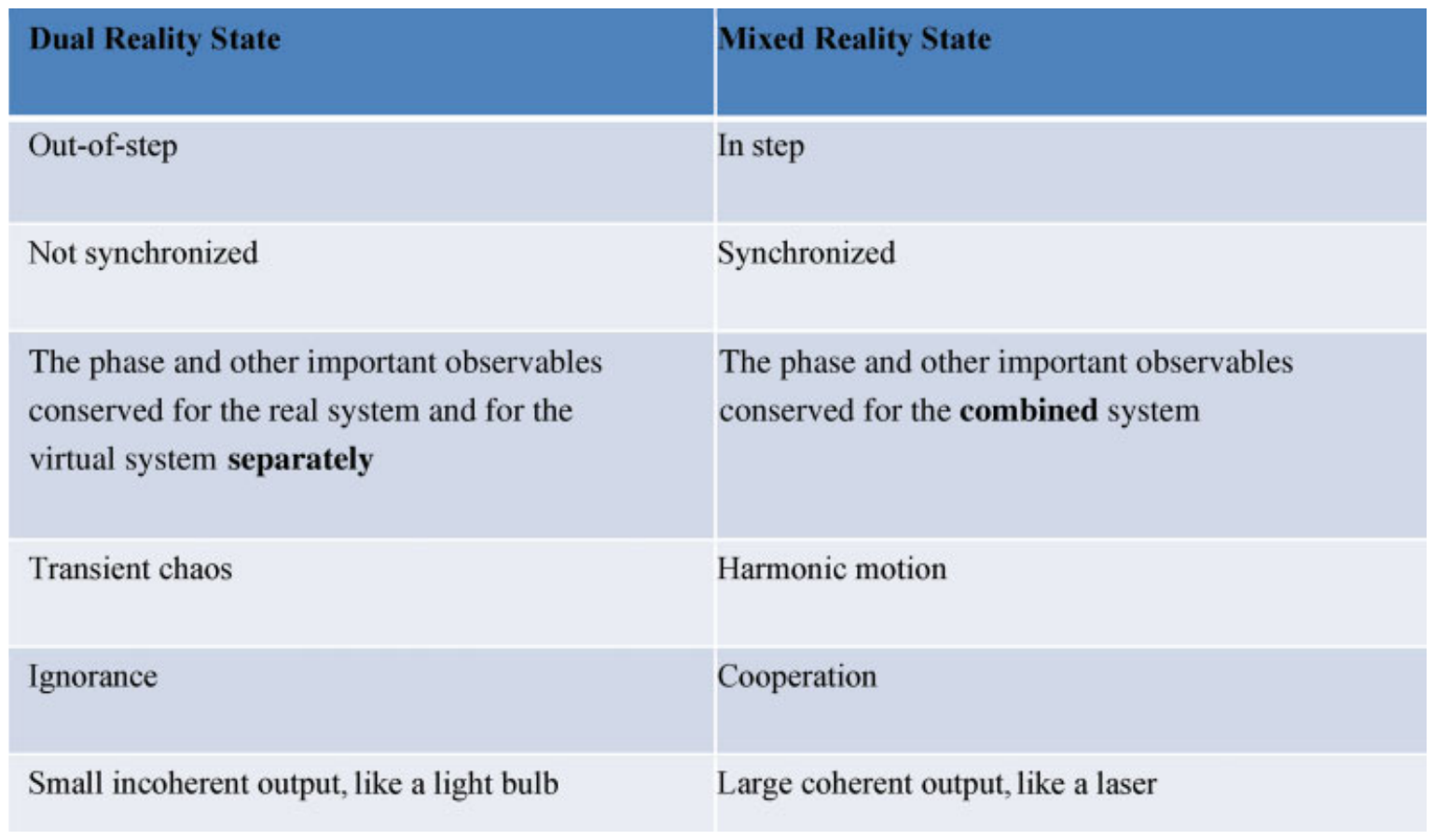}
      \caption{Properties of the motion of a system in a dual reality state and a mixed reality state.\label{fig1}}
  \end{center}
      \end{figure}

In a dual reality state, (i) the motion of the real system and the motion of the virtual system are out of step and not synchronized; (ii) if the energy loss due to friction is negligible, the energy of the real system and the virtual system are conserved separately; (iii) the amplitude of the motion decreases; (iv) the real system does not strengthen the motion of the virtual system and vice versa; and (v) the interreality system cannot generate a large output signal (see Figure~\ref{fig1}).

Synchronization between two real systems has been studied extensively in the past~\cite{Strogatz_2003}. For example, in a laser the dynamics of the electrons of different molecules can synchronize and thus produce a laser pulse of high intensity. Synchronization between real systems can be exploited for desirable effects. Lasers are used in CD players, laser pointers, and as energy efficient light sources. However, synchronization can also lead to destruction and catastrophes. At the opening of the Millennium Bridge in London to pedestrian traffic, the bridge started to sway sideways, which in turn caused the pedestrians to walk in step with the motion of bridge and the other pedestrians. Eventually the bridge started to sway at a dangerously large amplitude and had to be closed. The \$32 million bridge reopened 2 years later, after \$8.9 million worth of additional dampers were installed~\cite{Ouellette_2008}.

The significance of the recent work is the discovery that synchronization can occur between a real and a virtual system, and that the resulting mixed reality state is separated from the dual reality state by a phase transition~\cite{Gintautas07PRE}.

Mixed reality states can be used for control and for system identification, including the control of the human
mind. Out-of-body experiences triggered by video feedback are probably the best recent example of an observed mixed reality state. In the experiment, two video cameras are pointed at a subject from behind, and the subject sees a video image of himself with 3D goggles~\cite{Miller_2007,Blakeslee_2007}. The experimenter uses two sticks, one strokes the person's chest for 2 min, while the second stick moves just under the camera lenses, as if it were touching the virtual body.
Then some of the subjects report an experience where they appear to see themselves from outside their bodies. In this situation, the real system is the body of the subject, and the virtual system is the video image of the person. Since the video image is updated on a timescale which is much faster than the motion of the subject, and since the subject would recognize any change of the video image much faster than any motion on the body of the video image, there is instantaneous bidirectional interaction between the video image and the subject, that is, the subject and the video image are an interreality system. Future versions of computer games such as Second Life may include a fast coupling to the real world with video cameras and probes, and may control the real world with computer-controlled
motors, actuators, and switches. We would expect that eventually mixed reality states of virtual environments and the real world suddenly emerge.

Because mixed reality states occur only when a virtual and a real system are sufficiently similar, we can use a virtual system to learn more about a real system. As we adjust the parameters of the virtual system to achieve mixed reality, we can develop estimates about the real system. An example is the Dynamic Neuron system. Typically one would measure the dynamics of a neuron and then systematically adjust the parameters of a model until a perfect match between the experimental dynamics and the experimental data is achieved. However, models of real neurons have numerous parameters, far too many for a systematic search and the experimental data
are often too noisy. To overcome this problem, Kullmann et al. coupled a real neuron to its virtual counterpart with an instantaneous interaction. When the real system and the virtual system synchronize, that is, enter into a mixed reality state, the model is assumed to be good and thereby certain parameters may be identified~\cite{Kullmann_2004}

\begin{figure}[htb]
    \begin{center}
      \includegraphics[width=0.98\columnwidth]{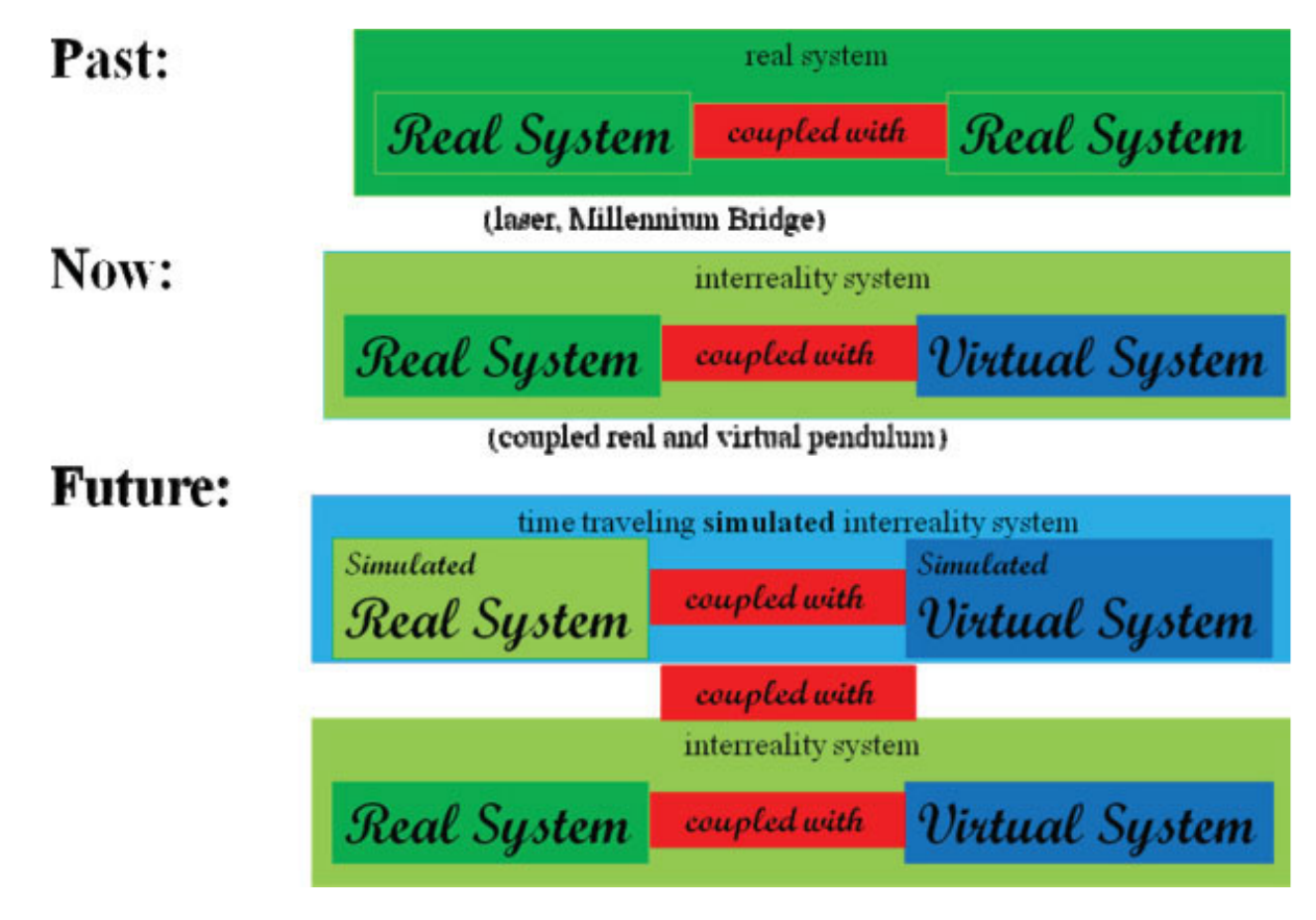}
      \caption{Synchronization in coupled real and virtual systems.\label{fig2}}
  \end{center}
      \end{figure}

When computers become even faster, it will be possible to couple an interreality system with its simulated counterpart. One of the unique advantages of the simulated interreality system is that it can fast-forward in time, or explore the past. In other words, the simulated interreality system can ``time travel'' (see Figure~\ref{fig2}). When an interreality system is coupled to its time - traveling simulated counterpart, mixed
reality states in the ``real'' interreality system can be stabilized. For instance, if a good driver is driving his car along the interstate, the driver has an image of the car, its surroundings, and its expected dynamical behavior in his mind, and uses the steering wheel and the other controls of the car to synchronize the motion of the car with its image. The wheels of car respond almost instantaneously to the turn of the steering wheel and the driver sees and feels the motion of the car quickly, hence the car and the driver are an interreality system. If the car does what the driver expects it to do, this may be considered a mixed reality state. If the car behaves unexpectedly, then the two may be considered a dual reality state. Usually the onset of the unexpected motion is sudden: there is a phase transition from a mixed reality state, where the car behaves as
expected, to a dual reality state, where the driver has lost control over the car.

Experienced drivers think ahead, that is, if they see difficult road conditions coming, they imagine themselves driving through the difficult area before they actually get to it~\cite{Davis_pc}. When they perform this mental rehearsal in preparation of an upcoming challenge, they fast-forward in time the imagined interreality system of themselves and their car. Just thinking about the upcoming hazard may cause them to slow down or change lanes right away, indicating that the simulated time-traveling interreality system is coupled
to the real interreality system. This coupling reduces the risk that the driver loses control over his car, that is, the coupling between the time-traveling imagined interreality system and the real interreality system is used to stabilize the mixed reality state of the driver and his car.

In the near future, computer hardware may become fast enough that it is not only possible to create mixed reality states, but to simulate the corresponding interreality system in real time or even have it time travel. The time-traveling simulated interreality system can be used to stabilize mixed reality states and make mixed reality states even more important.

This work was supported by the National Science Foundation through grant No. DMS 03-25939 ITR.


\begin{thebibliography}{7}
\expandafter\ifx\csname natexlab\endcsname\relax\def\natexlab#1{#1}\fi
\expandafter\ifx\csname bibnamefont\endcsname\relax
  \def\bibnamefont#1{#1}\fi
\expandafter\ifx\csname bibfnamefont\endcsname\relax
  \def\bibfnamefont#1{#1}\fi
\expandafter\ifx\csname citenamefont\endcsname\relax
  \def\citenamefont#1{#1}\fi
\expandafter\ifx\csname url\endcsname\relax
  \def\url#1{\texttt{#1}}\fi
\expandafter\ifx\csname urlprefix\endcsname\relax\def\urlprefix{URL }\fi
\providecommand{\bibinfo}[2]{#2}
\providecommand{\eprint}[2][]{\url{#2}}

\bibitem[{\citenamefont{Gintautas and H{\"u}bler}(2007)}]{Gintautas07PRE}
\bibinfo{author}{\bibfnamefont{V.}~\bibnamefont{Gintautas}} \bibnamefont{and}
  \bibinfo{author}{\bibfnamefont{A.~W.} \bibnamefont{H{\"u}bler}},
  \bibinfo{journal}{Phys.\ Rev.\ E} \textbf{\bibinfo{volume}{75}},
  \bibinfo{pages}{057201} (\bibinfo{year}{2007}).

\bibitem[{\citenamefont{Strogatz}(2003)}]{Strogatz_2003}
\bibinfo{author}{\bibfnamefont{S.}~\bibnamefont{Strogatz}},
  \emph{\bibinfo{title}{Sync: The Emerging Science of Spontaneous Order}}
  (\bibinfo{publisher}{Hyperion}, \bibinfo{address}{New York},
  \bibinfo{year}{2003}).

\bibitem[{\citenamefont{Ouellette}(2008)}]{Ouellette_2008}
\bibinfo{author}{\bibfnamefont{J.}~\bibnamefont{Ouellette}},
  \emph{\bibinfo{title}{Reality bites}}, \bibinfo{howpublished}{Retrieved from
  \url{http://twistedphysics.typepad.com/cocktail_party_physics/2008/03/reality-bites.html}}
  (\bibinfo{year}{2008}).

\bibitem[{\citenamefont{Miller}(2007)}]{Miller_2007}
\bibinfo{author}{\bibfnamefont{G.}~\bibnamefont{Miller}},
  \bibinfo{journal}{Science} \textbf{\bibinfo{volume}{317}},
  \bibinfo{pages}{1020} (\bibinfo{year}{2007}).

\bibitem[{\citenamefont{Blakeslee}(2007)}]{Blakeslee_2007}
\bibinfo{author}{\bibfnamefont{A.}~\bibnamefont{Blakeslee}},
  \bibinfo{journal}{The New York Times}  (\bibinfo{year}{2007}).

\bibitem[{\citenamefont{Kullmann et~al.}(2004)\citenamefont{Kullmann, Wheeler,
  Beacom, and Horn}}]{Kullmann_2004}
\bibinfo{author}{\bibfnamefont{P.~H.~M.} \bibnamefont{Kullmann}},
  \bibinfo{author}{\bibfnamefont{D.~W.} \bibnamefont{Wheeler}},
  \bibinfo{author}{\bibfnamefont{J.}~\bibnamefont{Beacom}}, \bibnamefont{and}
  \bibinfo{author}{\bibfnamefont{J.~P.} \bibnamefont{Horn}},
  \bibinfo{journal}{J.\ Neurophysiol.} \textbf{\bibinfo{volume}{91}},
  \bibinfo{pages}{542} (\bibinfo{year}{2004}).

\bibitem[{\citenamefont{Davis}()}]{Davis_pc}
\bibinfo{author}{\bibfnamefont{W.}~\bibnamefont{Davis}},
  \bibinfo{howpublished}{Private communication}.

\end{thebibliography}

\end{document}